# Statistics of the polariton condensate

Paolo Schwendimann and Antonio Quattropani

Institute of Theoretical Physics. Ecole Polytechnique Fédérale de Lausanne.

CH 1015 Lausanne-EPFL, Switzerland

The influence of polariton-polariton scattering on the statistics of the polariton condensate in a non-resonantly excited semiconductor microcavity is discussed. Taking advantage of the existence of a bottleneck in the exciton-polariton dispersion curve, the polariton states are separated into two domains: reservoir polaritons inside the bottleneck and active polaritons whose energy lies below the bottleneck. In the framework of the master equation formalism, the non-equilibrium stationary reduced density matrix is calculated and the statistics of polaritons in the condensate at $q=0$ is determined. The anomalous correlations between the polaritons in the condensate and the active ones are responsible for an enhancement of the noise in the condensate. As a consequence, the second order correlation function of the condensate doesn't show the full coherence that is characteristic of laser emission.

## I Introduction

The luminescent emission from a semiconductor microcavity that has been excited near the conduction band edge, has attracted much attention since the pioneering experiments of the Grenoble group[1]. These experiments show the insurgence of a strong emission in the exciton-polariton state with wave vector $q = 0$ after a threshold value for the pump field has been attained. The emission characteristics don't correspond to the ones of the usual semiconductor laser emission[2]. On the other hand, boson stimulation has been shown to occur in resonant polariton excitation experiments[3] that have been interpreted in terms of polariton parametric scattering [4-6]. Therefore, it is expected the experiments of [1] to be understood in the same framework as the resonant experiments, with one important difference. In fact, the onset of the polariton emission in [1, 2] has analogies with the onset of laser emission. In this case, the emission above threshold may be expected to show a high degree of coherence. Alternatively, since in the pump regime considered in the experiments, the polariton are assumed to obey Bose statistics, it has been conjectured that this new state may exhibit non-equilibrium polariton condensation at $q = 0$, the condensate being a non-equilibrium macroscopically populated state that doesn't correspond to a laser state. This interpretation relays on the fact that polaritons are mixed excitations of photons and excitons and for small wave vectors **q** their mass is exceedingly small thus allowing condensation to be observed at relatively high temperatures. In order to corroborate this second conjecture, several experiments have been performed that have allowed to gain much insight into the non-resonant emission [7-13] and eventually signatures of condensation in CdTe [14] and in GaAs [15] microcavities or laser-like action in CdTe [9] and in GaN [16] microcavities have been demonstrated. Notice that, in force of the finite lifetime of the polaritons (typically some picoseconds) and in the continuous pump configuration, the stationary state of the polaritons will not be a state of thermal equilibrium. As we have already pointed out above, the experimental results are interpreted according to two different pictures: non-equilibrium condensation of polaritons and polariton laser. In both pictures a macroscopically populated polariton state appears. Some information on which of these two pictures may better fit the experiments is obtained from the statistical properties of the emitted radiation. As it is well known, laser emission is characterized by a high coherence implying the normalized second order correlation function to have the value of one. Measurement of this same quantity for a polariton condensate presented in [10] indicate that no laser-like coherence is found. More recent experiments [17] show a peculiar behavior of the second order correlation that needs a detailed discussion in terms of the polariton dynamics but are not consistent with the conventional laser picture.

The theoretical approaches considered so far relay mainly on two different models: the model of non-linearly interacting polaritons [5, 18] and a generalization of the Dicke model [19]. In this paper we adopt the non-linear interacting polariton model that has been successful interpreting the non linear resonant polariton scattering [5]. Since the paper by Tassone and Yamamoto [20], much work has been done on the problem considered here either based on a Boltzmann equation approach [21, 22] or on a more refined kinetic approach that includes anomalous correlations [23]. These approaches, and in particular the one of [23] allow to enforce the interpretation of the emission as a signature of a non-equilibrium condensation, but don't allow to calculate the statistical properties of the emission. The approach of [24, 25] based on a master equation leads to the result that the statistics of the emission coincides with that of a laser, but relays on a oversimplified description of the polariton dynamics.

In this paper we present a quantum optical approach to the statistics of the emission. We consider a CdTe semiconductor microcavity that is excited near the conduction band edge in the continuous pump configuration. We take advantage of the existence of a bottleneck in the polariton dispersion curve in order to separate the polariton states into two domains. Polaritons inside the bottleneck are considered to act as a reservoir and the active polaritons whose energy lies below the bottleneck are considered to participate to the emission process. The stationary state resulting from the interplay between the polariton-polariton interaction below the bottleneck, the polariton-reservoir interaction and the cavity losses and other dissipation mechanism is not a state of thermal equilibrium. This stationary state is well described in the framework of the master equation formalism. In this approach we have access to the density matrix for the polaritons from which the polariton statistics is obtained. In particular, we show that the non-resonant, momentum conserving transition between the polaritons with wave vector $q=0$ and the ones with opposite wave vectors **q** and –**q** strongly influence the statistics of the mode with $q=0$, because they are responsible for anomalous correlations between this mode and the ones with **q** different from zero. These correlations act as a noise source for the mode with $q=0$ and prevent the emission showing full coherence above the emission threshold. The effect of non-resonant polariton-polariton scattering on the statistics had already been discussed in [26] using a simplified model in which only the non-resonant transitions between the reservoir modes and the polariton mode with $q=0$ were considered. Due to the absence of many polariton modes below the bottleneck, the relevance of the non-resonant scattering effects had been underestimated and a full coherent statistics was found above threshold. On the contrary, we show that high coherence is achieved at most in a very small region just above threshold in the system under study and that the statistics of the emission varies in a peculiar way in function of the excitation pump. This result, showing that the emission from this system is not comparable to the one of a conventional laser, is corroborated by recent experiments [17].

The paper is organized as follows. In Section II the system Hamiltonian is discussed and the basic equations are derived. In Section III the master equation describing the dynamics of the mode with $q = 0$ is derived. In Section IV the model describing the evolution of the modes wit $\mathbf{q} \neq 0$ under the influence of the pump is presented. Finally, in Section V the solutions of the master equation for the mode with $q = 0$ are presented and the statistics of the emission is discussed.

## II Basic equations

Our goal is to describe the photoluminescent emission from a system of interacting and non-resonantly excited polaritons in the stationary regime. The starting point is the Hamiltonian describing the interaction between exciton-polaritons [6]

$$H = \sum_{\mathbf{k}} \hbar \omega_{\mathbf{k}} P_{\mathbf{k}}^+ P_{\mathbf{k}} + \frac{1}{2} \sum_{\mathbf{k},\mathbf{k'},\mathbf{r}} V_{\mathbf{k},\mathbf{k'},\mathbf{r}} P_{\mathbf{k}-\mathbf{r}}^+ P_{\mathbf{k'}+\mathbf{r}}^+ P_{\mathbf{k}} P_{\mathbf{k'}} . \qquad (2.1a)$$

In $(2.1a)$ we have introduced the quantities

$$V_{\mathbf{k},\mathbf{k'},\mathbf{r}} = \frac{6 E_B \lambda_X^2}{A} X_{\mathbf{k}} X_{\mathbf{k'}} X_{\mathbf{k}} X_{\mathbf{k}+\mathbf{k'}-\mathbf{r}} + \frac{\hbar \Omega_R 16 \pi \lambda_X^2}{7A} X_{\mathbf{k}} X_{\mathbf{k'}} \left( |C_{\mathbf{k}}| X_{\mathbf{k'}+\mathbf{k}+\mathbf{r}} + |C_{\mathbf{k}+\mathbf{k'}-\mathbf{r}}| X_{\mathbf{k}} \right). \qquad (2.1b)$$

Here $A$ is the quantization area, $\lambda_X$ is the exciton radius, $E_B$ is the exciton binding energy and $\Omega_R$ is the Rabi frequency. The quantities $X_{\mathbf{k}}$ and $C_{\mathbf{k}}$ are the Hopfield coefficients of the exciton and the photon component of the polariton respectively. Following the approach of [21, 26] we take advantage of the presence of a bottleneck in the exciton polariton dispersion [27] and suppose that the polaritons whose wave vectors are inside the bottleneck region act as a reservoir. Since we are considering a continuous pump configuration, we assume the polaritons in the reservoir to be in a stationary state determined by the pump field. The photoluminescent emission is due to scattering processes between the bottleneck polaritons, denoted by an index $\mathbf{k}$ in the following, and the polaritons whose energy lies below the bottleneck energy, denoted by an index $\mathbf{q}$. The set of polaritons with index $\mathbf{q}$ is defined by the condition $q < q_{max}$, where we choose $q_{max}/k_0 \approx 0.1$, $k_0 = E_{ex}(0)/(\hbar v)$, $E_{ex}(0)$ is the exciton energy and v is the light velocity in the medium. The quantity $q_{max}$ represents the upper limit for the modulus of the active polariton wave vectors. We perform the separation between the reservoir polaritons and the active polaritons in (2.1) obtaining

$$\begin{aligned}
H_1 = & \sum_{\mathbf{q}=0}^{\mathbf{q}_{max}} \hbar\omega_{\mathbf{q}} P_{\mathbf{q}}^+ P_{\mathbf{q}} + \sum_{\mathbf{k}>\mathbf{q}_{max}} \hbar\omega_{\mathbf{k}} P_{\mathbf{k}}^+ P_{\mathbf{k}} + W_{0,0,0} P_0^+ P_0^+ P_0 P_0 + \sum_{\mathbf{q}\neq 0}^{\mathbf{q}_{max}} W_{0,\mathbf{q},-\mathbf{q}} \left( P_0^+ P_0^+ P_{\mathbf{q}} P_{-\mathbf{q}} + h.c. \right) + \\
& \sum_{\mathbf{k}>\mathbf{q}_{max}} W_{,0,\mathbf{k},-\mathbf{k}} \left( P_0^+ P_0^+ P_{\mathbf{k}} P_{-\mathbf{k}} + h.c. \right) + \sum_{\mathbf{q}\neq,\mathbf{q}'\mathbf{k},>\mathbf{q}_{max},} W_{\mathbf{k},\mathbf{q},\mathbf{q}'} (P_{\mathbf{q}}^+ P_{\mathbf{q}'}^+ P_{\mathbf{k}} P_{\mathbf{q}+\mathbf{q}'-\mathbf{k}} + h.c.) + \\
& \sum_{\mathbf{q}\neq,\mathbf{q}'\mathbf{k},>\mathbf{q}_{max},}^{\mathbf{q}_{max}} W_{\mathbf{k},\mathbf{q},\mathbf{q}'} (P_{\mathbf{q}}^+ P_{\mathbf{k}}^+ P_{\mathbf{q}} P_{\mathbf{q}-\mathbf{q}'+\mathbf{k}} + h.c.) + \\
& \sum_{\mathbf{q}=0,\mathbf{k},\mathbf{k}>\mathbf{q}_{max}}^{\mathbf{q}_{max}} W_{\mathbf{k},\mathbf{k}'\mathbf{q}} (P_{\mathbf{q}}^+ P_{\mathbf{k}+\mathbf{k}'-\mathbf{q}}^+ P_{\mathbf{k}} P_{\mathbf{k}'} + h.c.) + \\
& \sum_{\mathbf{q}=0,\mathbf{k}>\mathbf{q}_{max},,}^{\mathbf{q}_{max}} W_{\mathbf{q},\mathbf{k},\mathbf{q}} P_{\mathbf{k}}^+ P_{\mathbf{k}} P_{\mathbf{q}}^+ P_{\mathbf{q}} + \sum_{\mathbf{q},\mathbf{q}',\mathbf{r}} W_{\mathbf{q},\mathbf{q}'\mathbf{r}} P_{\mathbf{q}-\mathbf{r}}^+ P_{\mathbf{q}'+\mathbf{r}}^+ P_{\mathbf{q}} P_{\mathbf{q}'} .
\end{aligned} \qquad (2.2)$$

In (2.2) we have introduced the quantities

$$W_{\mathbf{r},\mathbf{r}',\mathbf{r}''} = \frac{1}{2}(V_{\mathbf{r},\mathbf{r}',\mathbf{r}''} + V_{\mathbf{r}',\mathbf{r},\mathbf{r}''}) .$$

The Hamiltonian (2.2) describes, besides the free evolution of the $\mathbf{q}$- and $\mathbf{k}$-polaritons, the following processes:

a) The scattering between the mode with $q = 0$ and both the opposite active modes $(\mathbf{q}, -\mathbf{q})$ and the reservoir modes $(\mathbf{k}, -\mathbf{k})$

b) The interaction between a mode $\mathbf{q}$ and the reservoir modes leading to damping and diffusion.

c) The resonant and non-resonant scattering between two different $\mathbf{q}$-modes and the reservoir.

d) The scattering between polariton numbers in the mode **q** and polariton numbers in the reservoir.

e) The scattering processes inside the mode with $q = 0$.

f) The scattering processes between the modes $\mathbf{q} \neq 0$.

The scattering processes (a) that conserve momentum but are non resonant play a central role in the emission process as shown in [26]. In particular, they are responsible for saturation and noise effects in the evolution of the mode with $q = 0$. As we shall see in the following, the contributions originating from these terms determine the moments of the polariton number distribution and lead to non-zero values of the anomalous correlations between the mode with $q = 0$ and the ones wit **q** different from zero. These processes will be related to a depletion of the mode with $q = 0$ in favor of the population of the modes with $\mathbf{q} \neq 0$ [23]. The same effects lead also to a decreasing of the coherence of the emission in the mode with $q = 0$. We conclude this short discussion of (2.2) noticing that in the following we shall not consider the contributions of the process (f) as well as of scattering processes inside the reservoir, because they are negligible in the framework of the approximations on which the present approach is based. We don't consider the contribution of polariton-phonon scattering in this approach, because in the approximation scheme considered here their contribution to the dissipation and injection rates are small compared to the ones originating in polariton-polariton scattering.

In order to obtain a description of the dynamics involving the modes **q** alone, we derive a master equation following the steps outlined in [26]. We introduce the projector $P$ defined as $P\rho = \rho_{\{\mathbf{k}\}}^{stat} Tr_{\{\mathbf{k}\}}\rho$, where $\rho_{\{\mathbf{k}\}}^{stat}$ is the stationary density operator of the reservoir modes alone. Starting from the Liouville-von Neumann equation for the total density operator $\rho$ and using the projector $P$ [28] we derive an equation for the quantity $\rho_M = Tr_{\{\mathbf{k}\}}\rho$ [26] that after having performed Born and Markov approximation reads

$$\hbar\frac{d}{dt}\rho_M(t) = M_0\rho_M(t) + \sum_{\mathbf{q}\neq 0}^{\mathbf{q}_{max}}\Lambda_\mathbf{q}\rho_M(t) + \sum_{\mathbf{q'},\mathbf{q}\neq\mathbf{q'}}^{\mathbf{q}_{max}}\left(\Lambda_{\mathbf{q},\mathbf{q'}1}\rho_M(t) + \Lambda_{\mathbf{q},\mathbf{q'}2}\rho_M(t)\right) -$$
$$i\sum_{\mathbf{q}\neq 0}^{\mathbf{q}_{max}}W_{0,\mathbf{q},-\mathbf{q}}\left[\left(P_\mathbf{q}^+P_{-\mathbf{q}}^+P_0P_0 + h.c.\right),\rho_M(t)\right] +$$
$$\sum_{\mathbf{q}\neq 0}^{\mathbf{q}_{max}}\gamma_\mathbf{q}\left(\left[P_\mathbf{q}\rho_M(t),P_\mathbf{q}^+\right] + \left[P_\mathbf{q},\rho_M(t)P_\mathbf{q}^+\right]\right) - i\sum_{q\neq 0}^{q_{max}}\hbar\hat{\omega}_\mathbf{q}\left[P_\mathbf{q}^+P_\mathbf{q},\rho_M(t)\right]. \qquad (2.3)$$

The energy $\hbar\hat{\omega}_\mathbf{q}$ consists of the energy of the free polaritons corrected by the shifts $\Delta\omega_\mathbf{q}$, whose explicit expressions are given in Appendix A. The explicit expressions for the different coefficients and of the operators $\Lambda_\mathbf{q}$ and $\Lambda_{\mathbf{q},\mathbf{q'}1,2}$ that appear in (2.3) are also given in Appendix A. Furthermore, in (2.3) we have introduced

$$\Lambda_0\rho_M(t) = -i\hbar\hat{\omega}_0\left[P_0^+P_0,\rho_0(t)\right] - i\left[W_{,0,0,0}P_0^{+2}P_0^2,\rho_M(t)\right] + \gamma_0\left(\left[P_0\rho_0(t),P_0^+\right] + h.c.\right)$$
$$+ \Gamma_0\left(\left[P_0\rho_0(t),P_0^+\right] + h.c.\right) + \Delta_0\left(\left[P_0^+\rho_0(t),P_0\right] + h.c\right)$$
$$+ \Gamma_1\left(\left[P_0^2\rho_M(t),P_0^{2+}\right] + \left[P_0^2,\rho_M(t)P_0^{2+}\right]\right)$$
$$+ \Delta_1\left(\left[P_0^{+2}\rho_M(t),P_0^2\right] + \left[P_0^{+2},\rho_M(t)P_0^2\right]\right). \qquad (2.4)$$

The expression (2.4) that describes the contribution of the interactions between the mode with $q=0$ and the reservoir modes in the maser equation coincides with Eq. (2.8) of Ref [26] ($\Gamma_0$ and $\Delta_0$ are identical with $\Gamma_2$ and $\Delta_2$ in [26]). The solution of (2.3) is a very difficult task because the whole space of the **q**-vectors has to be considered. Since we are interested in the emission at $q=0$, we reduce the number of the degrees of liberty of the system considering the reduced density operator $\rho_0 = Tr_\mathbf{q}\rho_M$. In order to obtain the equation for the evolution of $\rho_0$, we trace over all wave vectors $\mathbf{q}$ the master equation (2.3) obtaining

$$\hbar \frac{d}{dt}\rho_0(t) = \Lambda_0 \rho_0(t) - i \sum_{\mathbf{q} \neq 0}^{\mathbf{q}_{max}} W_{0\mathbf{q},-\mathbf{q}} \left( \left[ P_0^2, \ll P_{\mathbf{q}}^+ P_{-\mathbf{q}}^+ \gg \right] + \left[ P_0^{+2}, \ll P_{\mathbf{q}} P_{-\mathbf{q}} \gg \right] \right) +$$

$$\sum_{\mathbf{q} \neq 0}^{\mathbf{q}_{max}} \Gamma_{\mathbf{q}01} \left( \left[ P_0 \ll P_{\mathbf{q}} P_{\mathbf{q}}^+ \gg, P_0^+ \right] + h.c. \right) +$$

$$\sum_{\mathbf{q} \neq 0}^{\mathbf{q}_{max}} \Delta_{\mathbf{q}01} \left( \left[ P_0^+ \ll P_{\mathbf{q}}^+ P_{\mathbf{q}} \gg, P_0 \right] + h.c \right) \tag{2.5}$$

In (2.5) we have introduced the double-bracketed quantities $\ll A \gg = Tr_{\{\mathbf{q}\}} \rho_M A$ that depend on the polariton operators with $q = 0$ only. We have not written the terms with coefficients $\Gamma_{\mathbf{pq}2}, \Delta_{\mathbf{pq}2}$ in equation (2.5) because they are excessively small and will be neglected in the following. An analogous equation for the density operator $\rho_{\mathbf{p},-\mathbf{p}} = Tr_{0,\{\mathbf{q} \neq \mathbf{p}, -\mathbf{p}\}} \rho_M$ is derived in Appendix B. The equation (2.5) contains commutation operators acting on the quantities $\ll P_{\mathbf{q}}^+ P_{-\mathbf{q}}^+ \gg$, and $\ll P_{\mathbf{q}} P_{-\mathbf{q}} \gg$ that originate in the non-resonant scattering between the modes with $q = 0$ and the ones with $\mathbf{q} \neq 0$. The relevance of these terms is best illustrated when going over to the equation for the population of the mode $q = 0$. In fact, from (2.5) we derive the equation

$$\hbar \frac{d}{dt} \langle P_0^+ P_0 \rangle = -2(\Gamma_0 - \Delta_0 - 4\Delta_1 + \gamma_0) \langle P_0^+ P_0 \rangle + 8\Delta_1 + 2\Delta_0 -$$

$$2 \sum_{\mathbf{q} \neq 0}^{\mathbf{q}_{max}} \left( \Gamma_{0\mathbf{q}1} \langle (P_{\mathbf{q}}^+ P_{\mathbf{q}} + 1) \rangle \langle P_0^+ P_0 \rangle - \Delta_{0\mathbf{q}1} \langle P_0^+ P_0 + 1 \rangle \langle P_{\mathbf{q}}^+ P_{\mathbf{q}} \rangle \right) -$$

$$4(\Gamma_1 - \Delta_1) \langle P_0^+ P_0^+ P_0 P_0 \rangle + 4 \operatorname{Im} \sum_{\mathbf{q} \neq 0}^{\mathbf{q}_{max}} \left( W_{0\mathbf{q},-\mathbf{q}} < P_0^+ P_0^+ P_{\mathbf{q}} P_{-\mathbf{q}} > \right). \tag{2.6}$$

The imaginary part of the anomalous correlation $< P_0^+ P_0^+ P_{\mathbf{q}} P_{-\mathbf{q}} >$ is responsible for the coupling between the modes with $q = 0$ and the one with $\mathbf{q} \neq 0$ in (2.6) and originates in the terms containing the quantities $<<P_\mathbf{q} P_{-\mathbf{q}}>>$, and $<<P_\mathbf{q}^+ P_{-\mathbf{q}}^+>>$ in (2.5). As a consequence, the mode coupling manifests itself in the dynamics of the polariton system through the anomalous correlation. The role of the anomalous correlations in the theory of polariton condensation is also carefully discussed in [23].

## III The equation for the reduced density operator

In order to obtain a closed equation for $\rho_0$ we need to have an equation that relates the quantities $<<P_\mathbf{q}^+ P_{-\mathbf{q}}^+>>$, $<<P_\mathbf{q} P_{-\mathbf{q}}>>$ and the reduced density operator $\rho_0$. The quantity $<<P_\mathbf{q} P_{-\mathbf{q}}>>$ obeys the equation

$$\hbar \frac{d}{dt} <<P_\mathbf{q} P_{-\mathbf{q}}>> = \left(-2(i\hbar\omega_\mathbf{q} + \Gamma_{\mathbf{q}T}) + \Lambda_0\right) <<P_\mathbf{q} P_{-\mathbf{q}}> -$$

$$i \sum_{\mathbf{q}' \neq 0}^{\mathbf{q}_{max}} W_{0\mathbf{q}',-\mathbf{q}'} \left( \left[P_0^2, <<P_\mathbf{q}^+ P_{-\mathbf{q}}^+ P_\mathbf{q} P_{-\mathbf{q}} >>\right] + h.c. \right) -$$

$$i \sum_{\mathbf{q}' \neq 0}^{\mathbf{q}_{max}} W_{0\mathbf{q}',-\mathbf{q}'} \left[ P_0^2 \left( <<P_\mathbf{q}^+ P_\mathbf{q}>> + <<P_{-\mathbf{q}}^+ P_{-\mathbf{q}}>> + \rho_0 \right) + h.c. \right] +$$

$$\sum_{\mathbf{q}' \neq 0}^{\mathbf{q}_{max}} \Gamma_{\mathbf{q},\mathbf{q}',1} \left( \left[P_0^+ <<P_{\mathbf{q}'}^+ P_{-\mathbf{q}} P_\mathbf{q} P_{-\mathbf{q}'} >>, P_0\right] + h.c. \right) +$$

$$\sum_{\mathbf{q}' \neq 0}^{\mathbf{q}_{max}} \Delta_{\mathbf{q},\mathbf{q}',1} \left( \left[P_0 <<P_\mathbf{q} P_{-\mathbf{q}} P_\mathbf{q} P_{-\mathbf{q}'}^+ >>, P_0^+\right] + h.c. \right), \quad (3.1)$$

where we have introduced
$$\Gamma_{\mathbf{q}T} = \Gamma_\mathbf{q} + \gamma_\mathbf{q} - \Delta_\mathbf{q} . \quad (3.2)$$

A similar equation holds for $<<P_\mathbf{q}^+ P_{-\mathbf{q}}^+>>$. Equation (3.1) contains the density operator $\rho_0$ explicitly. We show that under suitable approximations (3.1) leads to a linear relation between $\rho_0$ and the quantity $<<P_\mathbf{q} P_{-\mathbf{q}}>>$. To this end, we formally integrate (3.1) in time, perform a Born approximation with respect to the operator $\Lambda_0$ and the Markov approximation. This last approximation is justified, because we are interested in stationary solutions i.e. solutions valid for times larger than the polariton relaxation times. Details are given in Appendix B. In order to obtain a close equation for $\rho_0$, we introduce the following factorization approximation for the density matrix

$$\rho_M = \rho_0 \otimes \rho_{\{\mathbf{q}\}} , \qquad (3.3)$$

where $\rho_0$ and $\rho_{\{\mathbf{q}\}}$ are the density operators obeying the master equations (2.5) and (B4). Furthermore, we assume that the dynamics of the modes $\mathbf{q} \neq 0$ follows a Boltzmann dynamics, i.e.

$$Tr_{\{\mathbf{q},0\}}\left(\rho_M P_{\mathbf{q}}^{+n} P_{\mathbf{q}'}^{m}\right) = \left\langle P_{\mathbf{q}}^{+n} P_{\mathbf{q}'}^{m}\right\rangle = \delta_{n,m}\delta_{\mathbf{q},\mathbf{q}'}\left\langle P_{\mathbf{q}}^{+} P_{\mathbf{q}}\right\rangle^n \qquad (3.4a)$$

and

$$Tr_{\{\mathbf{q},0\}}\left(\rho_M P_{\mathbf{q}}^{n} P_{\mathbf{q}'}^{n}\right) = Tr_{\{\mathbf{q},0\}}\left(\rho_M P_{\mathbf{q}}^{+n} P_{\mathbf{q}'}^{+n}\right) = 0 , \qquad (3.4b)$$

which implies the factorization of $\rho_{\{\mathbf{q}\}}$. Approximations (3.3) and (3.4) lead to the following relations

$$<< P_{\mathbf{q}}^{+} P_{\mathbf{q}} >> = \rho_0 < P_{\mathbf{q}}^{+} P_{\mathbf{q}} > \qquad (3.5a)$$

and

$$<< P_{\mathbf{q}} P_{-\mathbf{q}} >> = i \sum_{\mathbf{q}' \neq 0}^{\mathbf{q}_{max}} G_{\mathbf{q}}^r W_{0\mathbf{q}',-\mathbf{q}'}\left(\left[P_0^2, < P_{\mathbf{q}'}^{+} P_{\mathbf{q}'} >< P_{-\mathbf{q}'}^{+} P_{-\mathbf{q}'} > \rho_0\right] + h.c.\right) + \qquad (3.5b)$$

$$i \sum_{\mathbf{q}' \neq 0}^{\mathbf{q}_{max}} G_{\mathbf{q}}^r W_{0\mathbf{q}',-\mathbf{q}'}\left(P_0^2 \left(< P_{\mathbf{q}'}^{+} P_{\mathbf{q}'} > + < P_{-\mathbf{q}'}^{+} P_{-\mathbf{q}'} > +1\right)\rho_0 + h.c.\right)$$

We remark that (3.5b) represents a correction to the factorization approximation (3.3), as indicated in the Appendix B. Inserting (3.6) into (2.5), we obtain the closed equation for $\rho_0$

$$\hbar \frac{d}{dt}\rho_0(t) = -i\left[W_{0,0,0} P_0^{+2} P_0^2, \rho_0(t)\right] - i\left[\hbar\hat{\omega}_0 P_0^{+} P_0, \rho_0(t)\right] +$$
$$\Gamma_{0,TOT}\left(\left[P_0 \rho_0(t), P_0^{+}\right] + h.c.\right) + \Delta_{0,TOT}\left(\left[P_0^{+} \rho_0(t), P_0\right] + h.c.\right) +$$
$$\Gamma_{11}\left(\left[P_0^2 \rho_0(t), P_0^{+2}\right] + h.c.\right) + \Delta_{11}\left(\left[P_0^{+2} \rho_0(t), P_0^2\right] + h.c.\right) +$$
$$\gamma_0 \left(\left[P_0 \rho_0(t), P_0^{+}\right] + h.c.\right). \qquad (3.6)$$

The coefficients of the operator terms in (3.6) are defined as

$$\Gamma_{0,TOT} \equiv \left(\Gamma_0 + \sum_{\mathbf{q}\neq 0}^{\mathbf{q}_{max}} \Gamma_{0\mathbf{q}1} <P_\mathbf{q}^+ P_\mathbf{q}>\right), \quad (3.7a)$$

$$\Delta_{0,TOT} \equiv \left(\Delta_0 + \sum_{\mathbf{q}\neq 0}^{\mathbf{q}_{max}} \Delta_{0\mathbf{q}1} <P_\mathbf{q}^+ P_\mathbf{q}>\right), \quad (3.7b)$$

$$\Gamma_{11} \equiv \Gamma_1 + 2\sum_{\mathbf{q}\neq 0}^{\mathbf{q}_{max}} G_\mathbf{q}^r W_{0,\mathbf{q},-\mathbf{q}}^2 \left(<P_\mathbf{q}^+ P_\mathbf{q}>+1\right)\left(<P_{-\mathbf{q}}^+ P_{-\mathbf{q}}>+1\right), \quad (3.7c)$$

$$\Delta_{11} \equiv \Delta_1 + 2\sum_{\mathbf{q}\neq 0}^{\mathbf{q}_{max}} G_\mathbf{q}^r W_{0,\mathbf{q},-\mathbf{q}}^2 <P_\mathbf{q}^+ P_\mathbf{q}><P_{-\mathbf{q}}^+ P_{-\mathbf{q}}>, \quad (3.7d)$$

$$G_\mathbf{q}^r \equiv \mathrm{Re}\left(\frac{1}{i\hbar(\omega_\mathbf{q} - \omega_0) + \Gamma_{\mathbf{q}T} + \gamma_0}\right). \quad (3.7e)$$

The equation (3.6) is formally the same as (2.4), but with different coefficients given by (3.7). The dissipation rate $\Gamma_{0,TOT}$ and the injection rate $\Delta_{0,TOT}$ contain contributions related to the scattering between the modes with $\mathbf{q}\neq 0$. The new coefficients (3.7a) and (3.7b), modify the gain characteristics of the mode $q=0$. The new expressions for the coefficients of the saturation terms (3.7c) and (3.7d) influence the noise characteristics of the emission and thus its coherence properties and substantially modify $\Gamma_1$ and $\Delta_1$ [26] because these coefficients depend on the polariton population of the modes with $\mathbf{q}\neq 0$ as well as on their gain profile. Finally, from equation (3.1) in the stationary regime and using (3.3) we obtain an approximated expression for the anomalous correlation, namely

$$\langle P_0^+ P_0^+ P_\mathbf{q} P_{-\mathbf{q}}\rangle = iG_{\mathbf{q}0} W_{0,\mathbf{q}',-\mathbf{q}'} \left(\langle P_0^+ P_0\rangle+1\right)\langle P_\mathbf{q}^+ P_\mathbf{q}\rangle\langle P_{-\mathbf{q}}^+ P_{-\mathbf{q}}\rangle -$$
$$iG_{\mathbf{q}0} W_{0,\mathbf{q}',-\mathbf{q}'} \left(2\langle P_\mathbf{q}^+ P_\mathbf{q}\rangle+1\right)\langle P_0^{+2} P_0^2\rangle, \quad (3.8)$$

where

$$G_{\mathbf{q}0} \equiv \left(\Gamma_{\mathbf{q}T} + \Sigma_{TOT} + i\hbar(\omega_\mathbf{q} - \omega_0)\right)^{-1} \quad \text{and} \quad \Sigma_{TOT} \equiv -4\Delta_1 + \left(\Gamma_0 - \Delta_0 + \gamma_0\right).$$

The equation (3.8) shows the relation between the anomalous correlation and the second moment $\langle P_0^{+2} P_0^2 \rangle$ of the polariton distribution, whose value expresses the amount of noise in the polariton state. The master equation for $\rho_{\mathbf{p},-\mathbf{p}}$ is derived in closed form in the Appendix B following the same lines.

## IV The equations for the polariton population

In order to solve the master equation (3.6), we need to evaluate the stationary values of the population of the modes with $\mathbf{q} \neq 0$ as well as the dissipation and injection coefficients that appear in (3.6). We perform these calculations in the spirit of [21], which is based on the same separation between reservoir and active modes that leads to (2.2) and (2.3) and in which the polariton density and temperature in the reservoir are pump dependent quantities. Using (3.3), the equations describing the evolution of the polariton population are derived from the master equation (*B*7) and read

$$\hbar \frac{d}{dt} \langle P_{\mathbf{p}}^+ P_{\mathbf{p}} \rangle = -2\left(\Gamma_{\mathbf{p}} - \Delta_{\mathbf{p}} + \gamma_{\mathbf{p}} + \right)\langle P_{\mathbf{p}}^+ P_{\mathbf{p}} \rangle + 2\Delta_{\mathbf{p}} -$$

$$2\sum_{\mathbf{q} \neq p}^{\mathbf{q}_{\max}} \left( \Gamma_{\mathbf{p}\mathbf{q},1} \langle P_{\mathbf{p}}^+ P_{\mathbf{p}} \rangle \langle P_{\mathbf{q}} P_{\mathbf{q}}^+ \rangle - \Delta_{\mathbf{p}\mathbf{q},1} \langle P_{\mathbf{p}} P_{\mathbf{p}}^+ \rangle \langle P_{\mathbf{q}}^+ P_{\mathbf{q}} \rangle \right) +$$

$$W_{0,\mathbf{p},-\mathbf{p}}^2 \, 2\,\mathrm{Re}\left( \frac{<P_0^{+2} P_0^2>\left(\langle P_{\mathbf{p}}^+ P_{\mathbf{p}} \rangle + \langle P_{-\mathbf{p}}^+ P_{-\mathbf{p}} \rangle + 1\right)}{2\left(i\hbar\omega_0 + \Sigma_{TOT} + \gamma_{\mathbf{p}} - i\hbar\omega_{\mathbf{p}}\right)} \right) -$$

$$W_{0,\mathbf{p},-\mathbf{p}}^2 \, 4\,\mathrm{Re}\left( \frac{(\langle P_{\mathbf{p}}^+ P_{\mathbf{p}} \rangle)(\langle P_{-\mathbf{p}}^+ P_{-\mathbf{p}} \rangle + 1)(2<P_0^+ P_0>+1)}{2\left(i\hbar\omega_0 + \Sigma_{TOT} + \gamma_{\mathbf{p}} - i\hbar\omega_{\mathbf{p}}\right)} \right). \quad (4.1)$$

In order to solve the equations (4.1) we need the evolution equation for the number of polaritons with $q = 0$. This equation follows from (3.6) and reads

$$\hbar\frac{d}{dt}\langle P_0^+ P_0\rangle = -2(\Gamma_0 - \Delta_0 - 4\Delta_{11} + \gamma_0)\langle P_0^+ P_0\rangle + 8\Delta_{11} + 2\Delta_0 -$$

$$2\sum_{\mathbf{q}\neq 0}^{\mathbf{q}_{max}}\left(\Gamma_{0\mathbf{q}1}\langle(P_\mathbf{q}^+ P_\mathbf{q} +1)\rangle\langle P_0^+ P_0\rangle - \Delta_{0\mathbf{q}1}\langle P_0^+ P_0 +1\rangle\langle P_\mathbf{q}^+ P_\mathbf{q}\rangle\right)-$$

$$4(\Gamma_{11} - \Delta_{11})\langle P_0^+ P_0^+ P_0 P_0\rangle. \qquad (4.2)$$

The equation (4.2) is the first of a hierarchy that we need to truncate in order to obtain explicit solutions. In the following we shall adopt the factorization $\langle P_0^{+2} P_0^2\rangle = 2\langle P_0^+ P_0\rangle^2$ that insures consistency. This factorization guaranties that the values of $\langle P_0^+ P_0\rangle$ calculated both from (4.2) and from the numerical solution of (3.6) are of the same order of magnitude. The Eq. (4.1) and (4.2) generalize the ones given in [21]. Finally, we need calculate the injection rates $\Delta_\mathbf{q}, \Delta_{\mathbf{q},\mathbf{q}'1}$ and the dissipation rates $\Gamma_\mathbf{q}, \Gamma_{\mathbf{q},\mathbf{q}'1}$, that express the effects of the scattering with the reservoir polaritons. To this end, we have to specify the stationary state of the reservoir. In [26] this state had been chosen to be a thermal state with a fixed temperature and a pump dependent polariton density. In the present approach, following [21], we introduce a more flexible description in which the interaction with the external pump is explicitly included and the temperature of the reservoir varies as a consequence of scattering.

First of all we have to derive the stationary equation for the polariton density in the reservoir within the projector formalism used in the derivation of (2.3). We report only the equation that determines the stationary values of the population in the reservoir.

$$0 = \sum_{q=0}^{\mathbf{q}_{max}} \frac{1}{A}\left(\Gamma_\mathbf{q} + \sum_{q'=0}^{\mathbf{q}'_{max}} \Gamma_{\mathbf{q}\mathbf{q}'1}\left(\langle P_{\mathbf{q}'}^+ P_{\mathbf{q}'}\rangle + 1\right)\right)\langle P_\mathbf{q}^+ P_\mathbf{q}\rangle -$$

$$\sum_{q=0}^{\mathbf{q}_{max}} \frac{1}{A}\left(\Delta_\mathbf{q} + \sum_{q'=0}^{\mathbf{q}'_{max}} \Delta_{\mathbf{q}\mathbf{q}'1}\langle P_{\mathbf{q}'}^+ P_{\mathbf{q}'}\rangle\right)\left(\langle P_\mathbf{q}^+ P_\mathbf{q}\rangle + 1\right) - \frac{2\gamma_\mathbf{k}}{A}\langle P_\mathbf{k}^+ P_\mathbf{k}\rangle + F. \qquad (4.3)$$

Here $F$ is the pump amplitude that we have introduced phenomenological into (4.3) and the quantities $\langle P_0^+ P_0 \rangle$ and $\langle P_{\mathbf{q}}^+ P_{\mathbf{q}} \rangle$ are determined from (4.1) and (4.2). Finally we need explicit expressions for the reservoir population $\langle P_{\mathbf{k}}^+ P_{\mathbf{k}} \rangle$ that appear in the definitions (A1) to (A3). To this end, we introduced the following approximation [21]: we assume that the population of the reservoir modes is described by a Boltzmann distribution whose temperature and chemical potential are the same for all modes i.e.

$$\langle P_{\mathbf{k}}^+ P_{\mathbf{k}} \rangle^{stat} = N^{stat} \exp(-\hbar \omega_{\mathbf{k}} / k_B T_x), \tag{4.4a}$$
where
$$N^{stat} = n_x 2\pi \hbar^2 / (M k_B T_x). \tag{4.4b}$$

Here $\hbar \omega_{\mathbf{k}}$ is the energy of the polaritons, $k_B$ and $T_x$ are the Boltzmann constant and the reservoir temperature respectively, $M$ is the exciton mass and $n_x$ is the polariton number density. The temperature is determined from the stationary equation for the mean reservoir energy $k_B T_x$ [21]

$$0 = \sum_{q=0}^{q_{max}} \frac{1}{A} \left( \Gamma_{\mathbf{q}} + \sum_{q'=0}^{\mathbf{q}'_{max}} \Gamma_{\mathbf{q}\mathbf{q}'1} \left( \langle P_{\mathbf{q}'}^+ P_{\mathbf{q}'} \rangle + 1 \right) \right) \hbar \omega_{\mathbf{q}} \langle P_q^+ P_q \rangle -$$
$$\sum_{q=0}^{q_{max}} \frac{1}{A} \left( \Delta_{\mathbf{q}} + \sum_{q'=0}^{\mathbf{q}'_{max}} \Delta_{\mathbf{q}\mathbf{q}'1} \langle P_{\mathbf{q}'}^+ P_{\mathbf{q}'} \rangle \right) \hbar \omega_{\mathbf{q}} \left( \langle P_q^+ P_q \rangle + 1 \right) -$$
$$\left( 2\gamma_k k_B T_x n_x + \frac{AM}{\hbar^2} \gamma_{ph} (k_B^2 T_x^2 - k_B^2 T_L^2) n_x - F\, k_B T_L \right). \tag{4.5}$$

In (4.5), $T_L$ is the lattice temperature and $\gamma_{ph}$ is the linewidth of the phonons. The last term in (4.5) accounts for scattering between reservoir polaritons and acoustic phonons and has been introduced phenomenological following [21]. Both the temperature of the reservoir and the number density has to be determined. Introducing the Ansatz (4.4) in (4.3) and (4.5) and in particular in the coefficients $\Gamma$ and $\Delta$, and taking into account that in the Boltzmann regime $\langle P_{\mathbf{k}}^+ P_{\mathbf{k}} \rangle \ll 1$, we obtain a system of two coupled non-linear stationary equations for $n_x$ and $T_x$. However, solving such a system is a difficult task, therefore, we generalize these stationary equations to the time domain seeking for their stationary solutions. This step may be justified by going over to time-dependent projectors in the derivation of the master equation. We obtain the following equations:

$$\hbar \frac{d}{dt} n_x = n_x \sum_{q=0}^{q_{max}} \frac{1}{A} \left( X_{\mathbf{q}}(T_x) + \sum_{q'=0}^{\mathbf{q}'_{max}} X_{\mathbf{q},\mathbf{q}',1}(T_x) \left( \langle P_{\mathbf{q}'}^+ P_{\mathbf{q}'} \rangle + 1 \right) \right) \langle P_q^+ P_q \rangle -$$

$$\sum_{q=0}^{q_{max}} \frac{1}{A} \left( n_x^2 Y_{\mathbf{q}}(T_x) + n_x \sum_{q'=0}^{\mathbf{q}'_{max}} Z_{\mathbf{q},\mathbf{q}',1}(T_x) \langle P_{\mathbf{q}'}^+ P_{\mathbf{q}'} \rangle \right) \left( \langle P_q^+ P_q \rangle + 1 \right) - 2\gamma_k n_x + F \qquad (4.6a)$$

and

$$\frac{d}{dt} k_B T_x = n_x \sum_{q=0}^{q_{max}} \left( X_{\mathbf{q}}(T_x) + \sum_{q'=0}^{\mathbf{q}'_{max}} X_{\mathbf{q},\mathbf{q}',1}(T_x) \left( \langle P_{\mathbf{q}'}^+ P_{\mathbf{q}'} \rangle + 1 \right) \right) \hbar \omega_{\mathbf{q}} \langle P_q^+ P_q \rangle -$$

$$\sum_{q=0}^{q_{max}} \frac{1}{A} \left( n_x^2 Y_{\mathbf{q}}(T_x) + n_x \sum_{q'=0}^{\mathbf{q}'_{max}} Z_{\mathbf{q},\mathbf{q}',1}(T_x) \langle P_{\mathbf{q}'}^+ P_{\mathbf{q}'} \rangle \right) \hbar \omega_{\mathbf{q}} \left( \langle P_q^+ P_q \rangle + 1 \right) -$$

$$\left( 2\gamma_k k_B T_x n_x + \frac{AM}{\hbar^2} \gamma_{ph} (k_B^2 T_x^2 - k_B^2 T_L^2) n_x - F k_B T_L \right). \qquad (4.6b)$$

The quantities $X_{\mathbf{q}}$, $Y_{\mathbf{q}}$, $X_{\mathbf{q},\mathbf{q}',1}$, and $Z_{\mathbf{q},\mathbf{q}',1}$ are obtained from $\Gamma_{\mathbf{q}}$, $\Delta_{\mathbf{q}}$, $\Gamma_{\mathbf{q},\mathbf{q}'1}$, and $\Delta_{\mathbf{q},\mathbf{q}'1}$ by expressing the polariton population in the reservoir $\langle P_{\mathbf{k}}^+ P_{\mathbf{k}} \rangle$ through the Ansatz (4.4) and by putting $n_x$ in evidence, as shown in Appendix A The numerical solutions of the system of equations consisting of (4.1), (4.2), and (4.6) in the stationary limit are then used in order to calculate all coefficients that appear in (3.6). The system is solved using the following material parameters for a CdTe quantum well: $\hbar \omega_{q=0} = E_{exc}(0) - \hbar \Omega_R$, with $E_{exc}(0) = 1680 \, emV$, $2\hbar \Omega_R = 7 \, meV$, $\varepsilon = 7.4$, the total exciton mass $M = 0.296$, the exciton radius $\lambda_X = 47 \text{\AA}$ the quantization area $A = 6 \, 10^{-5} \, cm^2$. As an illustration we show in Fig.1 the population distribution as function of the wave vector component $q_x$ for different values of the pump intensity. From this figure it is clear that above threshold the emitted intensity, which is proportional to the polariton population, condenses at $q=0$. On the contrary below threshold the emission consists of two peaks centered on the wave vectors component $q_x$ and $-q_x$, indicating that in a three-dimensional plot it will be distributed along a circle of radius $q_x$.

**Fig. 1. The population distribution as a function of the wave vector q for different values of the pump intensity.**

## V Polariton statistics and conclusions

In order to discuss the statistical properties of the polaritons with $q = 0$, we need to solve the master equation (3.6). As it was already pointed out in [26], the diagonal and off-diagonal matrix elements of the density operator in (3.6) evolve separately. Since for t = 0 the off-diagonal elements of $\rho_0$ are zero, they vanish for any time. We concentrate on the stationary solution of the equation for the diagonal matrix elements of $\rho_0$ that reads

$$2(n+1)(n+2)\Gamma_{11}\rho_{S,n+2} + 2(n+1)(\Gamma_0 + \gamma_0)\rho_{S,n+1} -$$
$$\left[2(n+1)(n+2)\Delta_{11} + 2(n+1)\Delta_0 + 2n(\gamma_0 + \Gamma_0) + 2n(n-1)\Gamma_{11}\right]\rho_{S,n}$$
$$+2n\Delta_0\rho_{S,n-1} + 2n(n-1)\Delta_{11}\rho_{S,n-2} = 0 \:. \tag{5.1}$$

The solution of (5.1) is found numerically by the technique outlined in [26]. We use the material parameters for a CdTe quantum well already given at the end of the preceding section. The solution of (5.1) shows that a strong amplification sets on, as soon as the condition threshold given by

$$\left(\Delta_{0,TOT} + 4\Delta_{11} - \Gamma_{0,TOT} - \gamma_0\right) > 0, \tag{5.2}$$

is fulfilled. This last relation indicates that amplification requires the injection rate to be larger than the dissipation rate. Therefore, the population of the state with $q=0$ has to be different from zero, and in this case bosonic stimulation takes place. The probability distribution for the polariton population in function of the pump density is presented in Fig. 2. We notice, that the distribution changes its characteristics when the threshold value of the pump is reached such that the condition (5.2) is satisfied. Below threshold the polariton distribution corresponds to a geometrical distribution that is characteristic of incoherent emission. Above threshold the polariton number distribution vanishes for n = 0 and shows a maximum for n different from zero. For values of the pump slightly above threshold, the polariton distribution tends toward the symmetric distribution characteristic of a state with high coherence. However, for growing pump intensities the distributions becomes asymmetric and it's maximum only slightly shifts towards larger values of n. This fact indicates that the coherence of the polariton state is degraded by noise when the pump intensity grows. In order to verify this point, we present in Fig. 3 the second order normalized correlation at the initial time $g^{(2)}(0)$. From Fig. 3 follows that below threshold $g^{(2)}(0)$ has the value 2 that is predicted for incoherent processes. Above threshold $g^{(2)}(0)$ abruptly diminishes approaching the value of one, which is characteristic of a coherent laser source, but doesn't reach this value implying that full coherence is not achieved. Furthermore, for slightly higher value of the pump density, $g^{(2)}(0)$ starts growing again and the coherence of the polariton field is thus reduced. Finally for even higher values of the pump density $g^{(2)}(0)$ diminishes again showing a behavior comparable to the one presented in [10]. The behavior of $g^{(2)}(0)$ in Fig. 3 qualitatively reproduces very recent measurements performed in a CdTe microcavity [17]. This behavior of $g^{(2)}(0)$ is understood in terms of scattering processes between the polariton mode with $q = 0$ and the ones with $\mathbf{q} \neq 0$. The effects of these scattering processes enter in the master equation (3.6) through the sums on the occupation numbers of the polariton modes with $\mathbf{q} \neq 0$ that appear in the coefficients (3.7). As it will be shown below, these terms are responsible for saturation effects with growing pump density. We notice that, as already pointed out in [26], the fluctuations appear as source terms in the equations describing the evolution of the polariton number as well as of the quantity $g^{(2)}(0)$, as it follows from (4.2) and from the equation

$$\hbar\frac{d}{dt}\langle P_0^+ P_0^+ P_0 P_0 \rangle = (24\Delta_{11} + 4\Delta_{0,TOT} - 4(\Gamma_{11} + \Gamma_{0,TOT}) - 4\gamma_0)\langle P_0^+ P_0^+ P_0 P_0 \rangle -$$
$$8(\Gamma_{11} - \Delta_{11})\langle P_0^{+3} P_0^3 \rangle + (64\Delta_{11} + 8\Delta_{0,TOtT})\langle P_0^+ P_0 \rangle + 8\Delta_{11} \quad (5.3)$$

**Fig. 2. Probability distribution of the polaritons with $q=0$ for different values of the normalized pump intensity $F/F_S$. The threshold value of the pump is $F_S = 3.37$ meV$\mu$m$^{-2}$. Material parameters for CdTe are used.**
**(a)** $F/F_S = 0.967, 1.031, 1.095.$    **(b)** $F/F_S = 1.160, 1.224, 1.289.$

**Fig. 3.** Second order normalized correlation $g^{(2)}(0)$ for the *q=0* polaritons as a function of the normalized pump intensity $F/F_s$. $F_S = 3.37$ meV$\mu$m$^{-2}$. The dots indicate $g^{(2)}(0)$ without the effects of the scattering to the modes (q, -q). The vertical grey line indicates the position of the threshold. Material parameters for CdTe are used.

The source term in (5.3) originates in the non-resonant scattering processes and indicates that the fluctuations influence also the second moment of the polariton distribution in contrast to the laser case. We emphasize this point by comparing in Fig. 3 the correlation $g^{(2)}(0)$ calculated with and without the sums on the polariton population. Without the effects of scattering to the modes **q**, the second order correlation shows a coherence behavior comparable to the one of a conventional laser. On the contrary, including effects of the scattering processes reduces the interval in which higher coherence is observed to a small region near the threshold value of the pump density. We notice also that the higher order correlations $g^{(n)}(0)$ not only don't show complete coherence, as shown in Fig. 4, but their minimum value above threshold grows with the order of the correlation in contrast to the case of full coherence. The effect of the scattering between the modes $q = 0$ and $\mathbf{q} \neq 0$ appears also as a reduction of the number of polaritons emitted at $q = 0$ as shown in Fig. 5. The strong saturation that appears when the scattering processes between different polariton modes are included, indicates that population is transferred from the mode $q = 0$ to the other active modes. In [26] only scattering processes between $q = 0$ and the reservoir modes $\mathbf{k}$ and $-\mathbf{k}$ were considered thus underestimating the contributions of scattering to the noise above threshold. The population transfer between the modes originates in the anomalous correlations that appear in the dynamics of the system [23]. As already pointed out in Section II Eq. (2.6), the evolution of the polariton occupation at $q = 0$ is related to the one of the imaginary part of the anomalous correlation $< P_0^+ P_0^+ P_\mathbf{q} P_{-\mathbf{q}} >$. The contributions of the anomalous correlation to the master equation for $\rho_0$ are calculated in Section III, equations (3.1) to (3.5) and lead to the coefficients (3.7). The existence anomalous correlations between the different polariton modes is essential in order to obtain the correct behavior of the polariton statistics, as explicitly shown in (3.8). We finally remark that (5.3) differs from the corresponding equation for the one mode laser because of the presence of the source term $8\Delta_{11}$. We have also calculated the linewidth of the polariton emission that shows the same behavior as the one already presented in [26]. The linewidth dramatically decreases when approaching the threshold where it attains its minimum and starts once more growing above threshold, a behavior that has already been predicted in [29].

**Fig. 4. Third and fourth order normalized correlations $g^{(3)}(0)$, $g^{(4)}(0)$ for the *q=0* polaritons as a function of the normalized pump intensity $F/F_s$. $F_s = 3.37$ meV$\mu$m$^{-2}$. The dots indicate $g^{(n)}(0)$ without the effects of the scattering to the modes (q, -q). The vertical grey line indicates the position of the threshold. Material parameters for CdTe are used.the position of the threshold.**

**Fig. 5. Population of polaritons emitted at $q=0$ as a function of the normalized pump intensity $F/F_s$. $F_S = 3.37$ meV$\mu$m$^{-2}$. The dots indicate the population without the effects of the scattering to the modes (q, -q). The vertical grey line indicates the position of the threshold. Material parameters for CdTe are used.**

We conclude by noticing that the interpretation of the experiments in term of conventional laser emission is not compatible with the results presented above. Laser emission is characterized by $g^{(2)}(0)=1$ for a large range of values of the pump in contrast to the results shown above. This fact may be better understood when looking at the mechanisms that underlay laser emission and polariton emission. Laser emission is based on stimulated emission from a strongly populated level and the threshold condition is requires a population inversion to be achieved. On the contrary, in the polariton case the threshold condition is expressed by (5.2). This condition implies the injection rate into the polariton mode with $q=0$ to be larger than the losses. Therefore the population of the lowest polariton energy level has to be large enough to encompass the losses. The condition (5.2) is the reverse of the threshold condition for the conventional laser and indicates that final state stimulation is the relevant mechanism underlying the polariton emission process. Therefore, when the population of the lowest energy state diminishes in consequence of scattering to other higher energy levels, the characteristics of the emission change.

We resume the results obtained so far as follows. We have evaluated the stationary, out of equilibrium, macroscopically populated polariton state at $q=0$. This stationary state results from the interplay between the pump, the incoherent dissipation mechanisms (losses etc.) and the polariton-polariton scattering. The analysis of the statistics of the polaritons in this state shows that it is not a highly coherent state, as it would be expected from a conventional laser or from a fully condensed system, in which the condensed state is a coherent state. These characteristics are a consequence of polariton-polariton scattering that prevents the condensate to be completely filled up and introduces a strong noise component in its statistics. Our result appears to be consistent with recent measurements. We are in fact in presence of a new macroscopic quantum state out of equilibrium with peculiar statistical properties: the state is neither fully coherent nor incoherent. In the literature this state is assumed to correspond to that of a "polariton laser". Although the behavior of the polariton population resembles to the one of a laser, the name "polariton laser" is misleading because the emission under study doesn't show the coherence properties of laser radiation.

**Acknowledgments**: We acknowledge fruitful discussion with B. Deveaud, M. Richard, D. Sarchi and V.Savona.

## Appendix A

In this Appendix we introduce the definitions of the different operators and coefficients that appear in equation (2.3), which correspond to the contributions of the different interactions listed in Section II.

a) The interaction between a **q**-polariton and the reservoir leads to

$$\Lambda_{\mathbf{q}}\rho_M(t) = \Gamma_{\mathbf{q}}\left([P_{\mathbf{q}}\rho_M(t), P_{\mathbf{q}}^+] + [P_{\mathbf{q}}, \rho_M(t)P_{\mathbf{q}}^+]\right) - i\hbar\Delta\Omega(\mathbf{q})[P_{\mathbf{q}}^+ P_{\mathbf{q}}, \rho_M(t)]$$
$$\Delta_{\mathbf{q}}\left([P_{\mathbf{q}}^+\rho_M(t), P_{\mathbf{q}}] + [P_{\mathbf{q}}^+, \rho_M(t)P_{\mathbf{q}}]\right), \qquad (A1a)$$

$$\Gamma_{\mathbf{q}} = 2 \sum_{\mathbf{k},\mathbf{k'}>\mathbf{q}_{max},} \text{Re}\, G_{(\mathbf{k},\mathbf{k'},\mathbf{k}+\mathbf{k'}-\mathbf{q})+} \left|W_{\mathbf{k},\mathbf{k'q}}\right|^2 \times$$
$$\left[ \langle P^+_{\mathbf{k}+\mathbf{k'}-\mathbf{q}} P_{\mathbf{k}+\mathbf{k'}-\mathbf{q}} \rangle \left(\langle P^+_{\mathbf{k}} P_{\mathbf{k}} \rangle + 1\right)\left(\langle P^+_{\mathbf{k'}} P_{\mathbf{k'}} \rangle + 1\right) \right], \quad (A1b)$$

$$\hbar\Delta\Omega(\mathbf{q}) = 2 \sum_{\mathbf{k},\mathbf{k'}>\mathbf{q}_{max},} \text{Im}\, G_{(\mathbf{k},\mathbf{k'},\mathbf{k}+\mathbf{k'}-\mathbf{q})+} \left|W_{\mathbf{k},\mathbf{k'q}}\right|^2 \times$$
$$\left[ \langle P^+_{\mathbf{k}+\mathbf{k'}-\mathbf{q}} P_{\mathbf{k}+\mathbf{k'}-\mathbf{q}} \rangle \left(\langle P^+_{\mathbf{k}} P_{\mathbf{k}} \rangle + \langle P^+_{\mathbf{k'}} P_{\mathbf{k'}} \rangle + 1\right) - \langle P^+_{\mathbf{k}} P_{\mathbf{k}} \rangle \langle P^+_{\mathbf{k'}} P_{\mathbf{k'}} \rangle \right], \quad (A1c)$$

$$\Delta_{\mathbf{q}} = 2 \sum_{\mathbf{k},\mathbf{k'}>\mathbf{q}_{max},} \text{Re}\, G_{(\mathbf{k},\mathbf{k'},\mathbf{k}+\mathbf{k'}-\mathbf{q})+} \left|W_{\mathbf{k},\mathbf{k'q}}\right|^2 \times$$
$$\left[ \left(\langle P^+_{\mathbf{k}+\mathbf{k}-\mathbf{q'}} P_{\mathbf{k}+\mathbf{k'}-\mathbf{q}} \rangle + 1\right) \langle P^+_{\mathbf{k}} P_{\mathbf{k}} \rangle \langle P^+_{\mathbf{k'}} P_{\mathbf{k'}} \rangle \right]. \quad (A1d)$$

b) The two polariton interaction with the reservoir involving the annihilation/ of a **q**-polariton and the creation of a **q'**–polariton or vice versa leads to

$$\Lambda_{\mathbf{q},\mathbf{q'},1}\rho_M(t) = \Gamma_{\mathbf{q},\mathbf{q'},1} \left( \left[ P^+_{\mathbf{q'}} P_{\mathbf{q}} \rho_M(t), P^+_{\mathbf{q}} P_{\mathbf{q'}} \right] + \left[ P^+_{\mathbf{q'}} P_{\mathbf{q}}, \rho_M(t) P^+_{\mathbf{q}} P_{\mathbf{q'}} \right] \right) -$$
$$2i\hbar\Delta\Omega(\mathbf{q},\mathbf{q'},1)\left[ P^+_{\mathbf{q'}} P_{\mathbf{q'}} P_{\mathbf{q}} P^+_{\mathbf{q}}, \rho_M(t)) \right] +$$
$$\Delta_{\mathbf{q},\mathbf{q'},1} \left( \left[ P^+_{\mathbf{q}} P_{\mathbf{q'}} \rho_M(t), P^+_{\mathbf{q'}} P_{\mathbf{q}} \right] + \left[ P^+_{\mathbf{q}} P_{\mathbf{q'}}, \rho_M(t) P^+_{\mathbf{q'}} P_{\mathbf{q}} \right] \right), \quad (A2a)$$

$$\Gamma_{\mathbf{q},\mathbf{q'}1} = 2 \sum_{\mathbf{k'}>\mathbf{q}_{max},} \text{Re}\, G_{(\mathbf{q},\mathbf{q'k}+\mathbf{q'}-\mathbf{q})+} \left|W_{\mathbf{k},\mathbf{q},\mathbf{q'}}\right|^2 \left(\langle P^+_{\mathbf{k}+\mathbf{q'}-\mathbf{q}} P_{\mathbf{k}+\mathbf{q'}-\mathbf{q}} \rangle + 1\right)\langle P^+_{\mathbf{k}} P_{\mathbf{k}} \rangle, \quad (A2b)$$

$$\hbar\Delta\Omega(\mathbf{q},\mathbf{q'},1) = 2 \sum_{\mathbf{k'}>\mathbf{q}_{max},} \text{Im}\, G_{(\mathbf{q},\mathbf{q'k}+\mathbf{q'}-\mathbf{q})+} \left|W_{\mathbf{k},\mathbf{q},\mathbf{q'}}\right|^2 \left(\langle P^+_{\mathbf{k}+\mathbf{q}} P_{\mathbf{k}+\mathbf{q}} \rangle + \langle P^+_{\mathbf{k}} P_{\mathbf{k}} \rangle\right), \quad (A2c)$$

$$\Delta_{\mathbf{q},\mathbf{q'}1} = 2 \sum_{\mathbf{k'}>\mathbf{q}_{max},} \text{Re}\, G_{(\mathbf{q},\mathbf{q'k}+\mathbf{q'}-\mathbf{q})+} \left|W_{\mathbf{k},\mathbf{q},\mathbf{q'}}\right|^2 \langle P^+_{\mathbf{k}+\mathbf{q'}-\mathbf{q}} P_{\mathbf{k}+\mathbf{q'}-\mathbf{q}} \rangle \left(\langle P^+_{\mathbf{k}} P_{\mathbf{k}} \rangle + 1\right). \quad (A2d)$$

c) The two-polariton interaction with the reservoir involving the annihilation/creation of a **q**-polariton and a **q'** polariton leads to

$$\Lambda_{\mathbf{q},\mathbf{q'},2}\rho_M(t) = \Gamma_{\mathbf{q},\mathbf{q'},2} \left( \left[ P_{\mathbf{q}} P_{\mathbf{q'}} \rho_M(t), P^+_{\mathbf{q}} P^+_{\mathbf{q'}} \right] + \left[ P_{\mathbf{q}} P_{\mathbf{q'}}, \rho_M(t) P^+_{\mathbf{q}} P^+_{\mathbf{q'}} \right] \right) -$$
$$2i\hbar\Delta\Omega(\mathbf{q},\mathbf{q'},2)\left[ P^+_{\mathbf{q}} P^+_{\mathbf{q'}} P_{\mathbf{q}} P_{\mathbf{q'}}, \rho_M(t)) \right] +$$
$$\Delta_{\mathbf{q},\mathbf{q'},2} \left( \left[ P^+_{\mathbf{q}} P^+_{\mathbf{q'}} \rho_M(t), P_{\mathbf{q}} P_{\mathbf{q'}} \right] + \left[ P^+_{\mathbf{q}} P^+_{\mathbf{q'}}, \rho_M(t) P_{\mathbf{q}} P_{\mathbf{q'}} \right] \right), \quad (A5c)$$

$$\Gamma_{\mathbf{q}, \mathbf{q'}, 2} = \sum_{\mathbf{k} > \mathbf{q}_{max},} \operatorname{Re} G_{(\mathbf{q'}+,\mathbf{q}-\mathbf{k},\mathbf{q'},\mathbf{q})+} W_{\mathbf{k},\mathbf{q'},\mathbf{q}} W_{\mathbf{k},\mathbf{q'},\mathbf{q}} \left( \left\langle P^+_{\mathbf{q'}+,\mathbf{q}-\mathbf{k}} P_{\mathbf{q'}+,\mathbf{q}-\mathbf{k}} \right\rangle + 1 \right) \left( \left\langle P^+_{\mathbf{k}} P_{\mathbf{k}} \right\rangle + 1 \right), \quad (A3b)$$

$$\hbar \Delta \Omega(\mathbf{q}, \mathbf{q'}, 2) = \sum_{\mathbf{k'} > \mathbf{q}_{max},} \operatorname{Im} G_{(\mathbf{q'}+,\mathbf{q}-\mathbf{k},\mathbf{q'},\mathbf{q})+} W_{\mathbf{k},\mathbf{q'},\mathbf{q}} W_{\mathbf{k},\mathbf{q'},\mathbf{q}} \times$$

$$\left( \left\langle P^+_{\mathbf{q'}+,\mathbf{q}-\mathbf{k}} P_{\mathbf{q'}+,\mathbf{q}-\mathbf{k}} \right\rangle + \left\langle P^+_{\mathbf{k}} P_{\mathbf{k}} \right\rangle \right), \quad (A3c)$$

$$\Delta_{\mathbf{q}, \mathbf{q'}, 2} = 2 \sum_{\mathbf{k'} > \mathbf{q}_{max},} \operatorname{Re} G_{(\mathbf{q'}+,\mathbf{q}-\mathbf{k},\mathbf{q'},\mathbf{q})+} W_{\mathbf{k},\mathbf{q'},\mathbf{q}} W_{\mathbf{k},\mathbf{q'},\mathbf{q}} \left\langle P^+_{\mathbf{q'}+,\mathbf{q}-\mathbf{k}} P_{\mathbf{q'}+,\mathbf{q}-\mathbf{k}} \right\rangle \left\langle P^+_{\mathbf{k}} P_{\mathbf{k}} \right\rangle, \quad (A3d)$$

with

$$G_{(\mathbf{q'},\mathbf{q},\mathbf{q'}-\mathbf{q}+\mathbf{k})+} = \frac{-i\hbar(\omega_{\mathbf{q}} - \omega_{\mathbf{k}} + \omega_{\mathbf{q'}-\mathbf{q}+\mathbf{k}} - \omega_{\mathbf{q'}}) + (\gamma_{\mathbf{q'}} + \gamma_{\mathbf{k}} + \gamma_{\mathbf{q'}-\mathbf{q}+\mathbf{k}} + \gamma_{\mathbf{q}})}{\hbar^2 (\omega_{\mathbf{q}} - \omega_{\mathbf{k}} + \omega_{\mathbf{q'}-\mathbf{q}+\mathbf{k}} - \omega_{\mathbf{q'}})^2 + (\gamma_{\mathbf{q'}} + \gamma_{\mathbf{k}} + \gamma_{\mathbf{q'}-\mathbf{q}+\mathbf{k}} + \gamma_{\mathbf{q}})^2}, \quad (A4a)$$

$$G_{(\mathbf{q'},\mathbf{q},\mathbf{q'}+\mathbf{q}-\mathbf{k})+} = \frac{i\hbar(\omega_{\mathbf{k}} + \omega_{\mathbf{q'}+\mathbf{q}-\mathbf{k}} - \omega_{\mathbf{q'}} - \omega_{\mathbf{q}}) + (\gamma_{\mathbf{q'}} + \gamma_{\mathbf{k}} + \gamma_{\mathbf{q'}+\mathbf{q}-\mathbf{k}} + \gamma_{\mathbf{q}})}{\hbar^2 (\omega_{\mathbf{k}} + \omega_{\mathbf{q'}+\mathbf{q}-\mathbf{k}} - \omega_{\mathbf{q'}} - \omega_{\mathbf{q}})^2 + (\gamma_{\mathbf{q'}} + \gamma_{\mathbf{k}} + \gamma_{\mathbf{q'}+\mathbf{q}-\mathbf{k}} + \gamma_{\mathbf{q}})^2}, \quad (A4b)$$

$$G_{(\mathbf{k},\mathbf{k'},\mathbf{k}+\mathbf{k'}-\mathbf{q})+} = \frac{i\hbar(\omega_{\mathbf{q}} - \omega_{\mathbf{k}} - \omega_{\mathbf{k'}} + \omega_{\mathbf{k}+\mathbf{k'}-\mathbf{q}}) + (\gamma_{\mathbf{q}} + \gamma_{\mathbf{k}} + \gamma_{\mathbf{k'}} + \gamma_{\mathbf{k}+\mathbf{k'}-\mathbf{q}})}{\hbar^2 (\omega_{\mathbf{q}} - \omega_{\mathbf{k}} - \omega_{\mathbf{k'}} + \omega_{\mathbf{k}+\mathbf{k'}-\mathbf{q}})^2 + (\gamma_{\mathbf{q}} + \gamma_{\mathbf{k}} + \gamma_{\mathbf{k'}} + \gamma_{\mathbf{k}+\mathbf{k}-\mathbf{q}})^2}. \quad (A4c)$$

The frequency $\hat{\omega}_{\mathbf{q}}$ introduced in (2.3) is defined in terms of (A1c), (A2c), and (A3c) as

$$\hat{\omega}_{\mathbf{q}} = \omega_{\mathbf{q}} + \Delta \omega_{\mathbf{q}} = \omega_{\mathbf{q}} + \Delta \Omega(\mathbf{q}) + \sum_{\mathbf{q'}} \Delta \Omega(\mathbf{q}, \mathbf{q'}, 1) + \sum_{\mathbf{q'}} \Delta \Omega(\mathbf{q}, \mathbf{q'}, 2).$$

In Section IV we have introduced the Ansatz (4.4) that allows expressing the population $\left\langle P^+_{\mathbf{k}} P_{\mathbf{k}} \right\rangle$ as a function of the exciton density $n_x$ and of the temperature $T_x$. The explicit expressions for the different coefficients $\Gamma_q$, $\Delta_q$, $\Gamma_{q,q',1}$, and $\Delta_{q,q',1}$ that appear in (4.3) are obtained by introducing the approximation $\left\langle P^+_{\mathbf{k}} P_{\mathbf{k}} \right\rangle \ll 1$ that is fulfilled in thermal equilibrium, and by performing the replacement

$$\left\langle P^+_{\mathbf{k}} P_{\mathbf{k}} \right\rangle^{stat} = n_x \frac{2\pi \hbar^2}{M k_B T_x} \exp(-\hbar \omega_{\mathbf{k}} / k_B T_x) \quad (A5)$$

into the definitions (A3) to (A5). As an example we obtain

$$\Gamma_q \equiv n_x X_{\mathbf{q}}(T_x) = 2n_x \sum_{\mathbf{k},\mathbf{k'}>\mathbf{q}_{max},} \operatorname{Re} G_{(\mathbf{k},\mathbf{k'},\mathbf{k}+\mathbf{k'}-\mathbf{q})+} \left|W_{\mathbf{k},\mathbf{k'q}}\right|^2 \frac{2\pi\hbar^2}{Mk_B T_x} \exp(-\hbar\omega_{\mathbf{k}+\mathbf{k'}-\mathbf{q}}/k_B T_x) \quad (A6a)$$

and

$$\Delta_q \equiv n_x^2 Y_{\mathbf{q}}(T_x) = 2n_x^2 \sum_{\mathbf{k},\mathbf{k'}>\mathbf{q}_{max},} \operatorname{Re} G_{(\mathbf{k},\mathbf{k'},\mathbf{k}+\mathbf{k'}-\mathbf{q})+} \left|W_{\mathbf{k},\mathbf{k'q}}\right|^2 \times$$
$$\left(\frac{2\pi\hbar^2}{Mk_B T_x}\right)^2 \exp(-(\hbar\omega_{\mathbf{k}} - \hbar\omega_{\mathbf{k'}})/k_B T_x). \quad (A6b)$$

Analogous expressions follow for the remaining coefficients.

## Appendix B

In this Appendix we derive the equations that are presented in Section 3 as well as the master equation for $\rho_{\mathbf{q},-\mathbf{q}}$. We start by deriving the equations leading to (3.4). As already mentioned in Section 3, we integrate formally (3.1) in time obtaining

$$<<P_{\mathbf{q}} P_{-\mathbf{q}}>>(t) = \exp\left[\left(-2i\omega_{\mathbf{q}} - 2\Gamma_{\mathbf{q}T}/\hbar + \Lambda_0/\hbar\right)t\right] <<P_{\mathbf{q}} P_{-\mathbf{q}}>>(0)$$
$$+ \left[\left(-2i\omega_{\mathbf{q}} - 2\Gamma_{\mathbf{q}T}/\hbar + \Lambda_0/\hbar\right)t'\right] \Xi(t-t')dt', \quad (B1)$$

where

$$\Xi(t) = \sum_{\mathbf{q'}\neq 0}^{\mathbf{q}_{max}} \Gamma_{\mathbf{q},\mathbf{q'},1} \left(\left[P_0^+ << P_{\mathbf{q'}}^+ P_{-\mathbf{q}} P_{\mathbf{q}} P_{-\mathbf{q'}}>>, P_0\right] + h.c.\right) +$$
$$\sum_{\mathbf{q'}\neq 0}^{\mathbf{q}_{max}} \Delta_{\mathbf{q},\mathbf{q'},1} \left(\left[P_0 << P_{\mathbf{q}} P_{-\mathbf{q}} P_{\mathbf{q}} P_{-\mathbf{q'}}^+>>, P_0^+\right] + h.c.\right) -$$
$$i\sum_{\mathbf{q'}\neq 0}^{\mathbf{q}_{max}} W_{0\mathbf{q'},-\mathbf{q'}} \left(\left[P_0^2, << P_{\mathbf{q}}^+ P_{-\mathbf{q'}}^+ P_{\mathbf{q}} P_{-\mathbf{q}}>>\right] + h.c.\right) +$$
$$i\sum_{\mathbf{q'}\neq 0}^{\mathbf{q}_{max}} W_{0\mathbf{q'},-\mathbf{q'}} \left(P_0^2 \left(<<P_{\mathbf{q}}^+ P_{\mathbf{q}}>> + <<P_{-\mathbf{q}}^+ P_{-\mathbf{q}}>> + \rho_0\right) + h.c.\right).$$

We introduce Born approximation in the exponential operator in (B1) that consists in the following simplification

$$\Lambda_0 X \approx -i\hat{\omega}_0 \left[P_0^+ P_0, X\right] + \gamma_0 \left(\left[P_0 X, P_0^+\right] + h.c.\right). \quad (B2)$$

We also perform a long time approximation in the integral, because we are interested in stationary solutions and take advantage of the initial condition leading to $<< P_{\mathbf{q}} P_{-\mathbf{q}} >> (0) = 0$ and obtain

$$<< P_{\mathbf{q}} P_{-\mathbf{q}} >> = \sum_{\mathbf{q}' \neq 0}^{\mathbf{q}_{max}} G_{\mathbf{q}}^r \Gamma_{\mathbf{q},\mathbf{q}',1} \left( \left[ P_0^+ << P_{\mathbf{q}'}^+ P_{-\mathbf{q}} P_{\mathbf{q}} P_{-\mathbf{q}'} >>, P_0 \right] + h.c. \right) +$$

$$\sum_{\mathbf{q}' \neq 0}^{\mathbf{q}_{max}} G_{\mathbf{q}}^r \Delta_{\mathbf{q},\mathbf{q}',1} \left( \left[ P_0 << P_{\mathbf{q}'} P_{-\mathbf{q}} P_{\mathbf{q}} P_{-\mathbf{q}'}^+ >>, P_0^+ \right] + h.c. \right) -$$

$$i \sum_{\mathbf{q}' \neq 0}^{\mathbf{q}_{max}} G_{\mathbf{q}}^r W_{0\mathbf{q}',-\mathbf{q}'} \left( \left[ P_0^2, << P_{\mathbf{q}'}^+ P_{-\mathbf{q}'}^+ P_{\mathbf{q}} P_{-\mathbf{q}} >> \right] + h.c. \right) +$$

$$i \sum_{\mathbf{q}' \neq 0}^{\mathbf{q}_{max}} G_{\mathbf{q}}^r W_{0\mathbf{q}',-\mathbf{q}'} \left( P_0^2 \left( << P_{\mathbf{q}}^+ P_{\mathbf{q}} >> + << P_{-\mathbf{q}}^+ P_{-\mathbf{q}} >> + \rho_0 \right) + h.c \right), \qquad (B3)$$

where

$$G_{\mathbf{q}}^r = \text{Re}\left( \frac{1}{i\hbar(\omega_{\mathbf{q}} - \omega_0) + \Gamma_{\mathbf{q}T} + \gamma_0} \right).$$

An analogous equation holds for $<< P_{\mathbf{q}}^+ P_{-\mathbf{q}}^+ >>$. We now calculate $<< P_{\mathbf{q}} P_{-\mathbf{q}} >>$ by inserting the factorization (3.3) into the right hand side of (B3), which leads to (3.5b). The master equation for $\rho_{\mathbf{p},-\mathbf{p}}$ is obtained by taking the trace of (2.3) over all modes $\mathbf{q}$, including the mode $q = 0$, different from $(\mathbf{p},-\mathbf{p})$. By definition $\rho_{\mathbf{p},-\mathbf{p}} = \rho_{-\mathbf{p},\mathbf{p}}$, the result is

$$\hbar \frac{d}{dt} \rho_{\mathbf{p},-\mathbf{p}}(t) = \Pi(\mathbf{p},-\mathbf{p}) + \Pi(-\mathbf{p},\mathbf{p}) -$$
$$iW_{0,\mathbf{p},-\mathbf{p}}\left(\left[P_{\mathbf{p}}P_{-\mathbf{p}}, <<P_0^+ P_0^+>>\right] + \left[P_{\mathbf{p}}^+ P_{-\mathbf{p}}^+, <<P_0 P_0>>\right]\right) \quad (B4a)$$

with

$$\Pi(\mathbf{p},-\mathbf{p}) \equiv \Lambda_{\mathbf{p}}\rho_{\mathbf{p},-\mathbf{p}}(t) - i\hbar\hat{\omega}_{\mathbf{p}}\left[P_{\mathbf{p}}^+ P_{\mathbf{p}}, \rho_{\mathbf{p},-\mathbf{p}}(t)\right] + \gamma_{\mathbf{p}}\left(\left[P_{\mathbf{p}}\rho_{\mathbf{p},-\mathbf{p}}(t), P_{\mathbf{p}}^+\right] + h.c.\right) +$$

$$\sum_{\mathbf{q}\neq(\mathbf{p},-\mathbf{p})}^{\mathbf{q}_{max}} \Delta_{\mathbf{p},\mathbf{q},1}\left(\left[P_{\mathbf{p'}}^+ <<P_{\mathbf{q}}^+ P_{\mathbf{q}}>>, P_{\mathbf{p'}}\right] + h.c.\right) +$$

$$\sum_{\mathbf{q}\neq(\mathbf{p},-\mathbf{p})}^{\mathbf{q}_{max}} \Gamma_{\mathbf{p},\mathbf{q},1}\left(\left[P_{\mathbf{p}} <<P_{\mathbf{q}} P_{\mathbf{q}}^+>>, P_{\mathbf{p}}^+\right] + h.c.\right), \quad (B4b)$$

and

$$\Lambda_{\mathbf{p}}\rho_{\mathbf{p},-\mathbf{p}}(t) = \Gamma_{\mathbf{p}}\left(\left[P_{\mathbf{p}}\rho_{\mathbf{p},-\mathbf{p}}(t), P_{\mathbf{p}}^+\right] + h.c.\right) + \Delta_{\mathbf{p}}\left(\left[P_{\mathbf{p}}^+ \rho_{\mathbf{p},-\mathbf{p}}(t), P_{\mathbf{p}}\right] + h.c.\right). \quad (B4c)$$

Considerations similar to the ones leading from (B1) to (B3) allow obtaining a closed master equation for the density matrix of a generic polariton mode with $\mathbf{p} \neq \mathbf{0}$. From (2.3) we derive the equation for $<<P_0 P_0>>$ by taking the trace over the mode $q = 0$ alone obtaining

$$\hbar \frac{d}{dt} <<P_0 P_0>> = 2(-i\hbar\omega_0 - \Sigma_{TOT} + \Lambda_p) <<P_0 P_0>> -$$
$$2\left(\Gamma_1 - \Delta_1 - iW_{0,0,0}\right) <<P_0^+ P_0^3>> -$$
$$i\sum_{\mathbf{q}\neq\mathbf{0}}^{\mathbf{q}_{max}} W_{0\mathbf{q},-\mathbf{q}}\left(\left[P_{\mathbf{q}}^+ P_{-\mathbf{q}}^+, <<P_0^4>>\right] + \left[P_{\mathbf{q}} P_{-\mathbf{q}}, <<P_0^{+2} P_0^2>>\right]\right) -$$
$$2i\sum_{\mathbf{q}\neq\mathbf{0}}^{\mathbf{q}_{max}} W_{0\mathbf{q},-\mathbf{q}} P_{\mathbf{q}} P_{-\mathbf{q}}\left(2 <<P_0^+ P_0>> + \rho_{\mathbf{q}}\right) +$$
$$\sum_{\mathbf{q},\mathbf{q'}\neq 0}^{\mathbf{q}_{max}} \Gamma_{\mathbf{q},\mathbf{q'},1}\left(\left[P_{\mathbf{q}}^+ <<P_{\mathbf{q'}}^+ P_{\mathbf{q'}} P_0 P_0>>, P_{\mathbf{q}}\right] + h.c.\right) +$$
$$\sum_{\mathbf{q'}\neq 0}^{\mathbf{q}_{max}} \Delta_{\mathbf{q},\mathbf{q'},1}\left(\left[P_{\mathbf{q}} <<P_{\mathbf{q'}} P_{\mathbf{q'}}^+ P_0 P_0>>, P_{\mathbf{q}}^+\right] + h.c.\right), \quad (B5a)$$

where

$$\Sigma_{TOT} \equiv -4\Delta_1 + \left(\Gamma_0 - \Delta_0 + \gamma_0\right). \quad (B5b)$$

We use the procedure leading from (B1) to (3.4) and obtain

$$<< P_0 P_0 >> (t) = -\frac{i}{2\hbar\left(i\omega_0 + \Sigma_{TOT} + \gamma_p - i\omega_p\right)} \times$$

$$\left( \sum_{\mathbf{q}\neq\mathbf{0}}^{\mathbf{q}_{max}} W_{0\mathbf{q},-\mathbf{q}} \left[ P_{\mathbf{q}} P_{-\mathbf{q}}, << P_0^{+2} P_0^2 >> \right] - 2i \sum_{\mathbf{q}\neq\mathbf{0}}^{\mathbf{q}_{max}} W_{0\mathbf{q},-\mathbf{q}} P_{\mathbf{q}} P_{-\mathbf{q}} \left( 2 << P_0^+ P_0 >> + \rho_{\mathbf{q}} \right) \right). \quad (B6)$$

By taking the trace over $\mathbf{q} \neq (\mathbf{p},-\mathbf{p})$ in (B6) and by replacing it into (B1) we obtain the master equation

$$\hbar \frac{d}{dt} \rho_{\mathbf{p},-\mathbf{p}}(t) = \Pi(\mathbf{p},-\mathbf{p}) + \Pi(-\mathbf{p},+\mathbf{p}) +$$

$$\frac{1}{\hbar^2} \text{Re}\left( \frac{1}{\left(-i\omega_0 + \Sigma_{TOT} + \gamma_p + i\omega_p\right)} \right) |W_{0,\mathbf{p},-\mathbf{p}}|^2 \times$$

$$\left( \left[ P_{\mathbf{p}} P_{-\mathbf{p}} \left( 2\rho_{\mathbf{p},-\mathbf{p}} < P_0^+ P_0 > + \rho_{\mathbf{p},-\mathbf{p}} \right), P_{\mathbf{p}}^+ P_{-\mathbf{p}}^+ \right] \right) + h.c. +$$

$$\frac{1}{\hbar^2} \text{Re}\left( \frac{1}{\left(-i\omega_0 + \Sigma_{TOT} + \gamma_p + i\omega_p\right)} \right) |W_{0,\mathbf{p},-\mathbf{p}}|^2 \times$$

$$\left( \left[ \rho_{\mathbf{p},-\mathbf{p}} P_{\mathbf{p}} P_{-\mathbf{p}} < P_0^{+2} P_0^2 >, P_{\mathbf{p}}^+ P_{-\mathbf{p}}^+ \right] + h.c. \right) + h.c. \quad (B7)$$

Finally, we obtain (4.1) by multiplying (B7) with the polariton number operator $P_{\mathbf{p}}^+ P_{\mathbf{p}}$ and by taking the trace over the variables with the indexes $\mathbf{p}$ and $-\mathbf{p}$.

Figure Captions

Fig. 1. The population distribution as a function of the wave vector q for different values of the pump intensity.

Fig. 2. Probability distribution of the polaritons with $q=0$ for different values of the normalized pump intensity $F/F_s$. The threshold value of the pump is $F_S = 3.37$ meV$\mu$m$^{-2}$. Material parameters for CdTe are used.
(a) $F/F_S = 0.967, 1.031, 1.095$.
(b) $F/F_S = 1.160, 1.224, 1.289$.

Fig. 3. Second order normalized correlation $g^{(2)}(0)$ for the $q=0$ polaritons as a function of the normalized pump intensity $F/F_s$. $F_S = 3.37$ meV$\mu$m$^{-2}$. The dots indicate $g^{(2)}(0)$ without the effects of the scattering to the modes (q, -q). The vertical grey line indicates the position of the threshold. Material parameters for CdTe are used.

Fig. 4. Third and fourth order normalized correlations $g^{(3)}(0)$, $g^{(4)}(0)$ for the $q=0$ polaritons as a function of the normalized pump intensity $F/F_s$. $F_S = 3.37$ meV$\mu$m$^{-2}$. The dots indicate $g^{(n)}(0)$ without the effects of the scattering to the modes (q, -q). The vertical grey line indicates the position of the threshold. Material parameters for CdTe are used.

Fig. 5. Population of polaritons emitted at $q=0$ as a function of the normalized pump intensity $F/F_s$. $F_S = 3.37$ meV$\mu$m$^{-2}$. The dots indicate the population without the effects of the scattering to the modes (q, -q). The vertical grey line indicates the position of the threshold. Material parameters for CdTe are used.

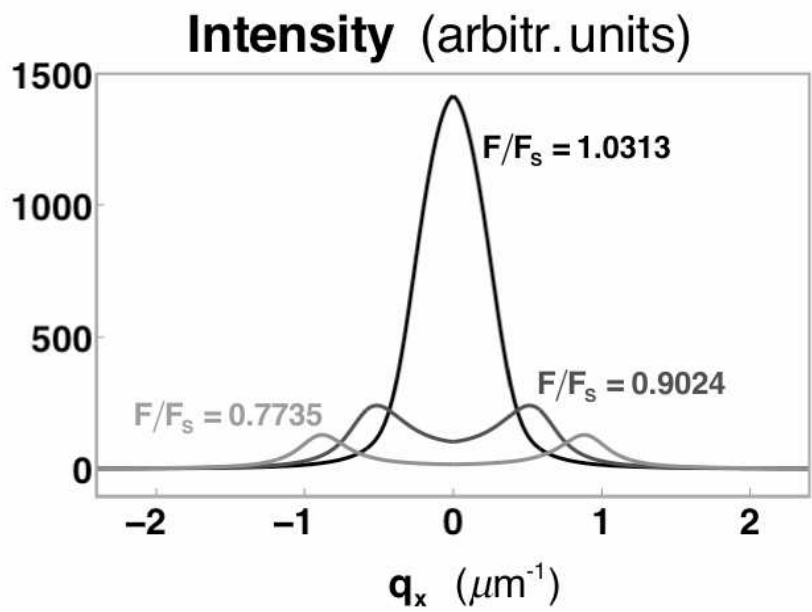

Figure 1

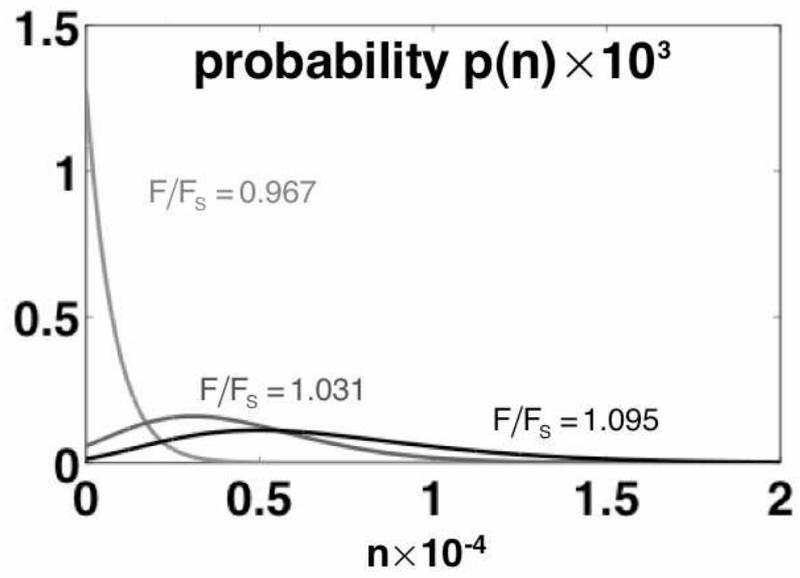

Figure 2a

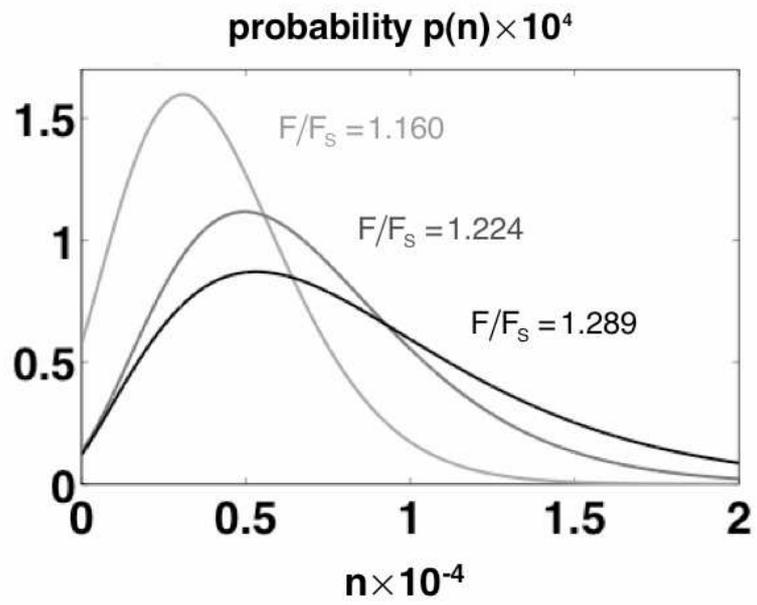

Figure 2b

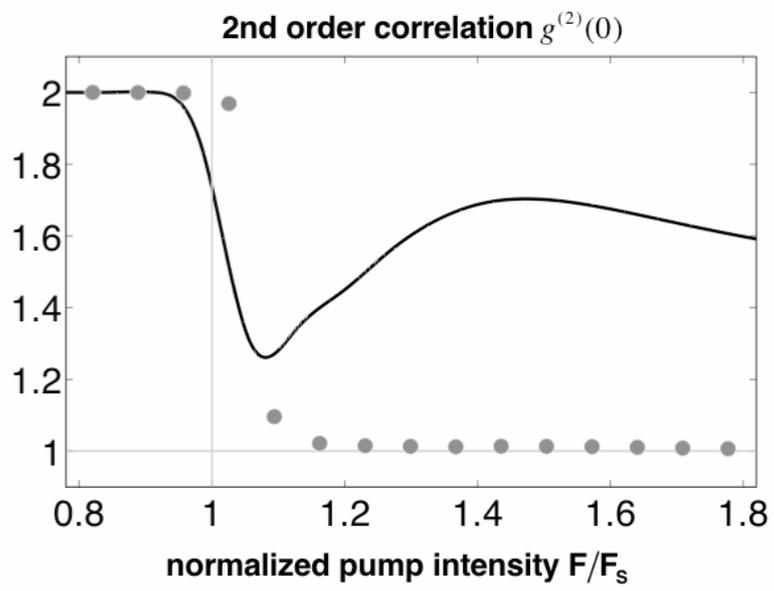

Figure 3

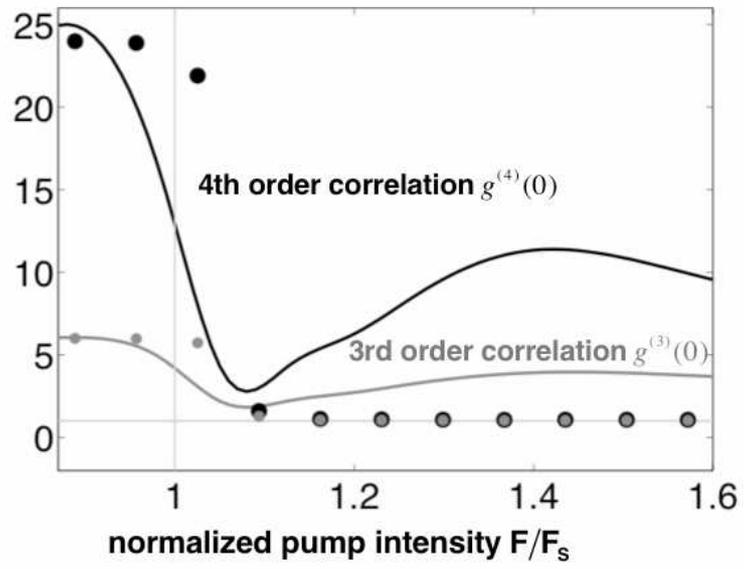

Figure 4

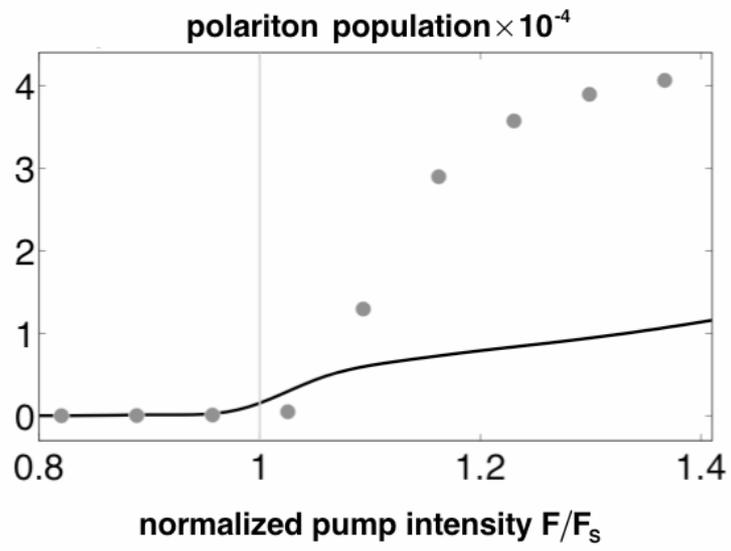

Figure 5